# The role of contacts in graphene transistors: A scanning photocurrent study

T. Mueller*, F. Xia, M. Freitag, J. Tsang, and Ph. Avouris*

*Email: tmuelle@us.ibm.com, avouris@us.ibm.com*

www.research.ibm.com/nanoscience

IBM Thomas J. Watson Research Center, Yorktown Heights, NY 10598

**Abstract.** A near-field scanning optical microscope is used to locally induce photocurrent in a graphene transistor with high spatial resolution. By analyzing the spatially resolved photo-response, we find that in the n-type conduction regime a p-n-p structure forms along the graphene device due to the doping of the graphene by the metal contacts. The modification of the electronic structure is not limited only underneath the metal electrodes, but extends 0.2–0.3 μm into the graphene channel. The asymmetric conduction behavior of electrons and holes that is commonly observed in graphene transistors is discussed in light of the potential profiles obtained from this photocurrent imaging approach. Furthermore, we show that photocurrent imaging can be used to probe single- / multi-layer graphene interfaces.

I. **INTRODUCTION**

Graphene, a single layer of graphite, is considered a promising material for use in future nanoelectronic devices [1]. The demonstration of current modulation by an electric field effect in graphene [2], followed by the recent demonstration of fast graphene transistors [3], has triggered extensive interest on the electrical properties and applications of this new material. Particularly, the unusual gate voltage dependence of the electrical conductivity (anomalous non-zero minimal conductivity [4-6] and differences in the conductances of electrons and holes [7-10]) is at the center of current interest. Most experiments to date probe the global response of a graphene transistor, i.e., they yield properties (for example the electrical conductance) averaged over the whole device. Similarly, in most simulations the graphene channel is treated as being homogeneous.

Recent experimental work [10-12], however, has provided evidence that charge inhomogeneity induced by the metal contacts might have a much stronger impact on the electrical transfer characteristics of graphene transistors than previously believed. To gain better understanding of the role of the contacts in graphene transistors, local characterization of the functioning devices using scanning probe techniques is clearly necessary. Scanning photocurrent (PC) microscopy has proven a useful tool for studying potential profiles in carbon nanotubes transistors [13-15], and recently also in graphene [16,17]. The resolution of a classical optical microscope, however, is restricted by diffraction to about half of the optical wavelength ($\sim\lambda/2$). An understanding of the photo-response, and hence potential profiles, on a smaller length-scale is desirable.

In this publication we report high-resolution PC imaging using near-field scanning optical microscopy (NSOM). NSOM overcomes the far-field resolution limit by bringing a light source of sub-wavelength size into close proximity ($<<\lambda$) to the sample surface. The resolution of the image is limited by the size of the probe aperture and not by the wavelength $\lambda$ of the light [18]. By analyzing the spatial variation of the PC in the vicinity of the metal contacts, we show that charge-transfer doping occurs underneath the contact metals and adjacent regions in the graphene channel, giving rise to asymmetric conduction characteristics for electrons and holes. In a complementary experiment, we also demonstrate charge transfer and photocurrent generation at single- / multi-layer graphene interfaces.

II. SAMPLE DESCRIPTION AND EXPERIMENTAL SETUP

The back-gated graphene transistors used in this study were prepared by mechanical exfoliation of Kish graphite using an adhesive tape and subsequent deposition of the flakes on a highly $p^+$-doped Si wafer, on which a 300 nm thick $SiO_2$ layer was grown by dry oxidation. Single layers of graphene were first identified visually using an optical microscope and further confirmed by Raman spectroscopy [19]. Source and drain Ti/Pd/Au (0.5/15/5 nm) electrodes were then deposited by electron-beam lithography, electron-beam evaporation of the metals, and lift-off. In a second lithography step, wide and thick Ti/Au (5/200 nm) bonding pads were patterned. The sample was then mounted in a ceramic chip carrier and wire bonds were made between the die and the package. In the as-prepared samples, the minimum conductance occurs at a back gate voltage of ~100 V. The samples were therefore annealed for several hours in an ultrahigh-vacuum chamber at 400 K. This procedure removes most of the doping adsorbates and water from the sample surface and shifts the Dirac point voltage close to 0 V. After taking out the samples from the vacuum chamber we typically observe that the Dirac point shifts back to 20-40 V, where it stays stable during the entire measurement process.

A commercial NSOM was adapted to carry out local PC measurements on the graphene devices. Fig. 1 shows the experimental setup and sample structure. Optical excitation is provided by a chopped (~1 kHz) Ar ion laser ($\lambda = 514.5$ nm). The laser light source is coupled into a metal-coated tapered optical fiber probe with a 100-nm aperture. The aperture locally illuminates the sample surface and the induced PC is recorded with a lock-in amplifier as the NSOM probe tip is scanned across the graphene transistor. The distance between the fiber tip and the sample is maintained at ~20 nm by applying a non-optical normal-force feedback technique. Taking into account the penetration of the light into the metal cladding of the NSOM probe and additional widening of the beam diameter by the tip-sample separation, we estimate an upper limit of the spatial resolution of ~150 nm. A topographic atomic force microscopy (AFM) image is acquired simultaneously with the PC image, allowing correlation of structural and PC properties at the same positions on the graphene transistor.

### III. PHOTOCURRENT NEAR THE METAL / GRAPHENE INTERFACE

On the left in Fig. 2, we show a scanning electron microscopy (SEM) image of one of our graphene devices, together with the electrical setup. PC measurements were performed under short-circuit conditions. The device exhibits the typical V-shaped conductance versus gate bias with a minimum at $V_G^{Dirac} = 40$ V, indicating a natural p-doping of the graphene (probably caused by trapped charge in the gate oxide). A mobility $\mu$ of approximately 0.1 m$^2$/Vs was extracted.

On the right in Fig. 2 we show a sequence of PC images of the device, taken at different gate biases $V_G$ between -60 and 100 V. From the topographic AFM image we are able to precisely determine the edges of the source and drain contacts (shown as dashed lines). The measurement at $V_G$ = -60 V displays strong PC ($I_{ph}$) with opposite polarities at the interfaces between graphene and the source and drain electrodes, respectively. As we increase the gate voltage, the PC gradually decreases, switches polarity, and increases again at larger positive voltages. By positioning the NSOM tip close to one of the contacts and sweeping the gate voltage, we determined the exact value of $V_G$ at which the sign of PC reverses: $V_G^{flat} \sim 20$ V. The presence of the strong PC spots close to the contact electrodes is due to the existence of local electric fields near the metal/graphene interfaces. For illumination in the middle of the device, the absence of a strong electric field will not separate the photo-excited electron-hole pairs and they will recombine rather efficiently. The overall trend of these findings is in line with what we have observed in far-field scanning PC measurements [17]. The high spatial resolution of the near-field technique, however, sheds light on various aspects of PC generation in graphene transistors that have not been revealed in previous work. Apart from the opposite polarity, the most striking differences between the p-type conduction regime and the n-type regime are (i) the spatial position of the PC maxima and (ii) the PC contribution from the metal contacts.

In order to extract quantitative information from the images, we plot in Figs. 3(a) and (b) the PC profiles at $V_G$ = -60 V and 100 V, respectively. The arrows in Fig. 2 mark the positions along which the profiles were taken. From Fig. 3(a) it is obvious that the PC at $V_G$ = -60 V is made up of two contributions. A strong and narrow response at the electrodes, that decays on a length scale of about 0.2 μm within the graphene sheet, and a much broader contribution from the contacts. At $V_G$ = 100 V [Fig. 3(b)], the PC has not only flipped its polarity, but has also moved ~0.28 μm away from the contacts, and has broadened to ~0.36 μm (full width at half maximum - FWHM). For another sample we obtain similar values [~0.22 μm and ~0.29 μm (FWHM), respectively]. PC from the metal contacts is now strongly suppressed.

$I_{ph}$ is a direct measure of the local potential gradient in the ~150 nm wide excitation region. In contrast to traditional semiconductors, the current resulting from carrier diffusion can be neglected in graphene because of the short lifetime $\tau$ of photo-excited carriers. Relaxation times $\tau$ of typically 0.1-2 ps have been reported [20, 21]. With these values and the mobility $\mu$ from above we estimate diffusion lengths $L_D = (V_t \mu \tau)^{1/2}$ ($V_t$ is the thermal voltage – 26 mV at room temperature) of ~15-70 nm, i.e. smaller than our excitation region [22]. A possible complication of the near-field PC imaging technique compared to traditional far-field microscopy is the presence of the metallized NSOM probe in close proximity to the graphene which could influence the potential in the transistor channel (and thus the PC generation) due to screening of the gate field. In order to minimize this impact we do not electrically ground the metal cladding of the NSOM tip, instead we let it float. Furthermore, we benefit from the fact that the graphene flake itself is a conductor that effectively shields the field produced by the gate.

The behavior of the PC discussed above can then be understood within a simple model that treats bending of the graphene bands as a result of charge transfer between the graphene sheet and the metal electrodes. Metals in contact with graphene pin the Fermi level below the electrodes and hence create a potential step within the graphene sheet [23]. As we will show, in our devices, the Pd contact introduces p-doping of the graphene underneath the electrodes. Thus, depending on the gate bias $V_G$, a p-p junction or a p-n junction forms in the vicinity of the electrode/graphene interface. In Figs. 4(a) and (b) we show band diagrams of the graphene transistor in the p-type conduction regime and the n-type regime, respectively. Since the PC is proportional to the potential gradient at the excitation position, we extract the band diagrams by numerical integration of the PC profiles. The energetic offset of the graphene bands with respect to the Fermi level is determined based on the following considerations: At zero applied gate bias ($V_G = 0$) we observe a weak $I_{ph}$ that is directed away from the source and drain electrodes, i.e., photo-excited electrons drift to the nearby electrode and holes toward the bulk of graphene. Since the minimum conductance for this device occurs at $V_G^{Dirac} = 40$ V, we may draw the band profile at $V_G = 0$ V as shown in Fig. 4(a) (dashed line). From a simple capacitor model we obtain an expression for the energetic difference between the Fermi level and the band edge in the bulk graphene channel as a function of applied gate bias [24]: $\Delta E = \hbar v_F \sqrt{\pi \alpha |V_G - V_G^{Dirac}|}$, where $\alpha = 7.2 \times 10^{10}$ cm$^{-2}$V$^{-1}$ and $\hbar v_F = 5.52$ eVÅ. With this equation we obtain $\Delta E(V_G = 0) \sim 0.17$ eV. The direction of the current flow requires the constant potential offset $\Delta \phi$ to be smaller than the potential in the center of the device: $\Delta \phi < \Delta E(V_G = 0)$. As the gate voltage is decreased to negative voltages, $\Delta E$ increases, whereas $\Delta \phi$ stays pinned at the contacts. The band bending at the contacts hence becomes steeper and $I_{ph}$ becomes stronger (solid line). When a positive gate voltage is applied the band first becomes flat, and eventually, the main body of the graphene becomes n-type when the band edge moves below the Fermi level [see Fig. 4(b)]. At flat-band condition ($V_G^{flat} \sim 20$ V) almost no PC is observed. This allows us to estimate the potential step at the graphene/electrode interface: $\Delta\phi = \hbar v_F \sqrt{\pi\alpha |V_G^{flat} - V_G^{Dirac}|} \sim 0.12$ eV. Because the contact region stays p-type even at positive gate biases, a p-n junction forms close to the electrode/graphene interface. Locally excited electron-hole pairs are separated in the strong electric field and contribute to PC. The most striking feature of Fig. 4(b) is that charge-transfer doping occurs not only underneath the electrodes, but extends hundreds of nanometers into adjacent regions in the graphene channel.

Within our model, we can also understand the PC response from the metal contacts. Carriers that are thermally excited in the metal contribute to the PC only if there exists an electric field at the electrode/graphene interface. This is obviously the case at negative gate voltages where the maximum of the electric field occurs right at the interface. Due to the direction of the field, only holes contribute to the current. At positive voltages, however, because of electrostatics, there is no significant band bending at the interface. Carriers have to diffuse through the field-free region before being separated by the strong electric field at the p-n junction. Given the rather short carrier diffusion length compared to the distance between the metal electrode and the p-n junction, most carriers recombine before they can reach the junction, resulting in suppressed PC. In addition to the strong response from the electrodes, we observe a weaker photo-response from the bulk of graphene. The existence of local electric fields is attributed to charged impurities in the substrate and residues of photoresist (PMMA) that cause random spatial variations of the local potential (electron-hole puddles) [25]. Note that this contribution is only observed in the vicinity of the Dirac point, whereas it vanishes at high negative and positive $V_G$. At high extrinsic carrier densities $n$ in graphene, the comparably small number of charged impurities introduces only small quantitative corrections. At low carrier densities, however, due to the low density-of-states, a small spatial variation of the local carrier concentration causes a strong variation of the local potential.

The contact doping discussed above causes an asymmetry between the p-n-p type conduction regime ($V_G < V_G^{Dirac}$) and the p-p-p type regime ($V_G > V_G^{Dirac}$) that is also reflected in electrical transport measurements [10,11]. The resistance that is associated with a p-n junction is larger than that of an equivalent p-p junction. This can easily be understood within a diffusive carrier transport model, where the resistance is simply obtained by integrating the local resistivity along the length of the junction [26, 27]. In the p-n case, the graphene band edge crosses the Fermi level and the carrier concentration in the junction hence approaches zero. This gives rise to an excess resistance with respect to the p-p case, where the Fermi level lies deep in the valence band. In the ballistic transport regime, the resistance of a p-n junction stems from the selective transmission of carriers, which only allows for the passage of particles that approach the junction in an almost perpendicular direction [28, 29]. The theory for ballistic propagation of carriers in a potential similar to that reported in Fig. 4(b) can be found in Ref. 30. Fogler *et al.* have introduced a dimensionless parameter $\beta = n' n_i^{-3/2}$ ($n'$ is the slope of the density profile at the Fermi level; $n_i$ is related to the mobility by $n_i = e/(\mu h)$, where $e$ is the elementary charge and *h* is Planck's constant) that separates the diffusive ($\beta \ll 1$) from the ballistic ($\beta \gg 1$) transport regime [31]. When calculating $\beta$ for our samples, we typically obtain values close to 1. Our samples are therefore in an intermediate regime, where the total resistance has diffusive and ballistic contributions. A detailed calculation is beyond the scope of this work and we refer the interested reader to the appropriate literature [30-32].

Experimentally, in our samples we do indeed observe the asymmetric conduction behavior for electrons and holes predicted in the previous paragraph. Following Ref. 10, we quantify this asymmetry by calculating the odd part of the device resistance $R_{odd} = [R(\Delta V_G) - R(-\Delta V_G)]/2$, where $\Delta V_G = V_G - V_G^{Dirac}$. We obtain a positive value which is consistent with our model [33]. For large $\Delta V_G$, the normalized resistance $R_{odd} \cdot W$ approaches a constant value of ~0.25 kΩ μm, independent of device length *L*. This is a clear indication that $R_{odd}$ is a contact resistance, rather than a resistance that is associated with different conductivities of electrons and holes in the graphene sheet [7]. In addition, our devices are approximately 10 times shorter than those in Ref. 7. The impact of the metal contacts on the conductance asymmetry is hence expected to dominate over the impact from the relatively short bulk graphene channel.

## IV. PHOTOCURRENT NEAR SINGLE- / MULTI-LAYER GRAPHENE INTERFACE

As discussed above, the charge transfer between the metal electrodes and the graphene sheet causes band bending near the metal/graphene interface. Band bending does also occur when a single layer of graphene is brought into contact with multi-layer graphene. Fig. 5 (a) shows the SEM image of a device that consists of a single-layer graphene (SLG) sheet (region "2") sandwiched between two sheets of multi-layer graphene (MLG) (regions "1" and "3"). The number of layers in the three regions was determined by Raman spectroscopy [19] and it was confirmed that region "2" is SLG, whereas regions "1" and "3" consist of two or possibly three layers.

In Fig. 5 (b) we show the PC image of the device recorded without applying a gate bias. In Fig. 5 (c) we plot the PC profile along the channel of the device. The dotted vertical lines represent the spatial positions of the metal electrodes and SLG / MLG interfaces, respectively. Apart from the strong and narrow PC in the vicinity of the metallic contact electrodes, we observe a weaker photoresponse with opposite polarities at the interfaces between SLG and MLG. The direction of the local electric field points from the SLG to the MLG sheet, as drawn schematically in Fig 5 (b). Upon local illumination, photo-generated carriers get separated and produce a PC in direction of the field. At present, we can only speculate on what causes the potential gradient at the interface between the two materials. It can, for example, be due to charge transfer between the different regions of the device. The SLG and BLG work functions can be tuned by applying a gate voltage, but also depend on surfaces dipoles imposed by adsorbates on top of the graphene surface and on the electronic structure of the material itself. As the two materials are brought into contact, the Fermi levels line up. As a result two charge layers are set up at the interface and an electric field is established. From the experimentally observed current flow direction from SLG to MLG one would conclude, that under experimental conditions similar to ours ($SiO_2$ substrate, ambient environment, hole-doping at zero gate bias), the measured work function of the multi-layer graphene is higher than that of the single layer. Other explanations, though, are possible, such as dipoles that are associated with the edges of the MLG [34]. Irrespective of what causes the electric field at the interface, our observation clearly demonstrates that a heterogeneous surface topography results in potential fluctuations and hence reduced carrier mobility in graphene.

**CONCLUSIONS**

We have studied the locally induced PC in graphene transistors by near-field optical excitation. We have shown that metal contacts have a strong impact on the electronic structure of the graphene channel and that this modification extends hundreds of nanometers away from the contacts. We have found that in the n-type conduction regime a p-n-p structure forms along the graphene channel due to Fermi level pinning in the graphene below the Ti/Pd contact electrodes. The existence of a p-n junction in the p-type conduction regimes gives rise to an excess resistance with respect to the n-type regime, resulting in an asymmetric conduction behavior for electrons and holes. Studies of interfaces between SLG and MLG have shown that a potential gradient occurs across the interface. The near-field PC spectroscopy method used here hence provides a powerful tool for the study of graphene-based electronic and optoelectronic devices.

**ACKNOWLEDGMENTS**

We would like to thank R. Golizadeh-Mojarad, Y. Lin, and V. Perebeinos for helpful discussions and B. Ek for technical assistance. T.M. acknowledges financial support by the Austrian Science Fund (FWF).

**Figure captions**

**Figure 1.** (Color online) Schematic illustration of the experimental setup and sample structure.

**Figure 2.** (Color) The left picture shows the SEM image of a graphene transistor and the electrical setup for PC measurements. On the right we show seven PC images taken at gate biases between -60 and +100 V. The dashed lines indicate the edges of the source and drain electrodes. The two scale bars on the bottom of the very right image are both 1 µm long.

**Figure 3.** (Color online) PC profiles at (a) $V_G$ = -60 V and (b) $V_G$ = 100 V along the arrows in Fig. 2. The dashed lines indicate the edges of the source and drain electrodes.

**Figure 4.** (Color online) (a) Band diagrams at $V_G$ = 0 V (dashed line) and $V_G$ = -60 V (solid line), obtained by numerical integration of the PC profiles in Fig. 3. $\Delta\phi$ describes the pinning of the Fermi level. Arrows indicate the flow of electrons and holes. (b) Band diagram at $V_G$ = 100 V, obtained by numerical integration of the PC profile in Fig. 3. It shows the formation of a p-n-p structure. The distance $d$ between the PC peaks is smaller than the device length $L$. There is no PC contribution from the contacts because the electric field at the electrode/graphene interface is nearly zero.

**Figure 5.** (Color) (a) SEM image of the device. Region "2" consists of single-layer graphene, regions "1" and "3" are multi-layer. (b) PC image recorded in the p-type conduction regime. The black dashed lines indicate the edges of the source and drain electrodes. The white dotted lines mark the interfaces between SLG and MLG. (c) PC profile along the channel of the device. The red line indicates the PC that is generated at the graphene interfaces.

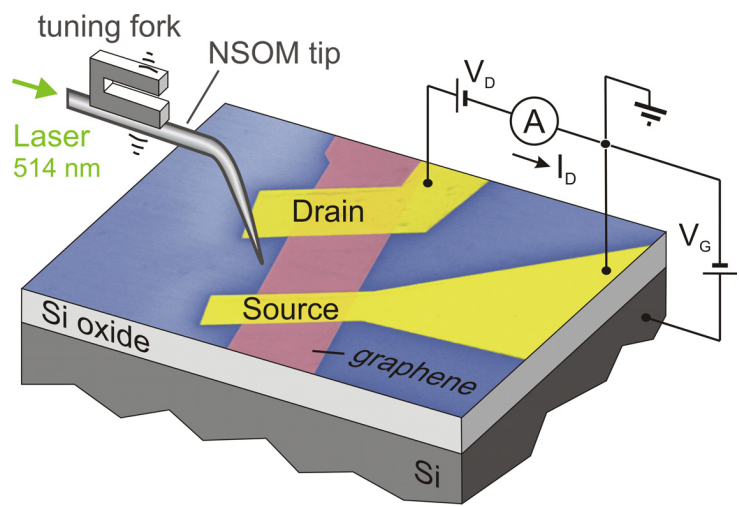

Figure 1, Mueller et al.

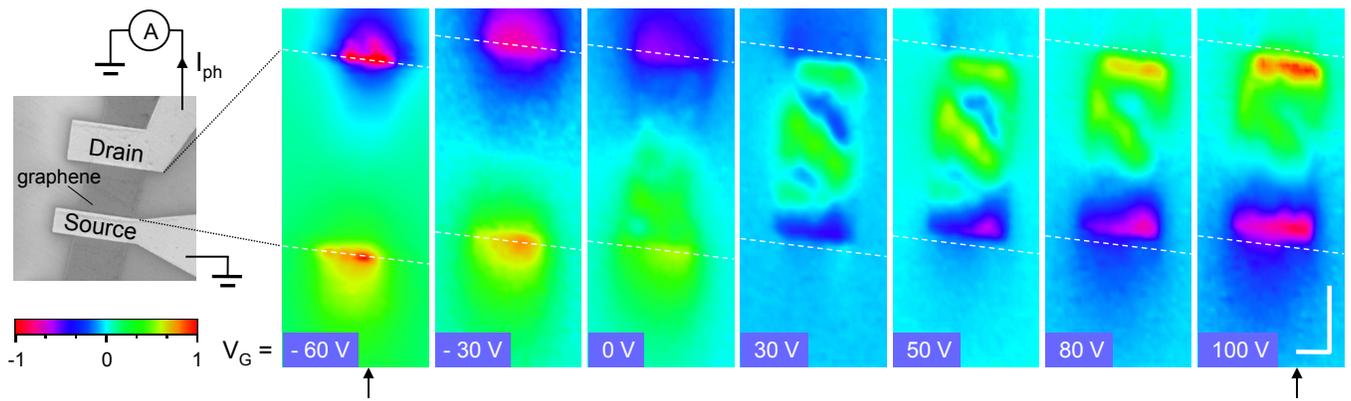

Figure 2, Mueller et al.

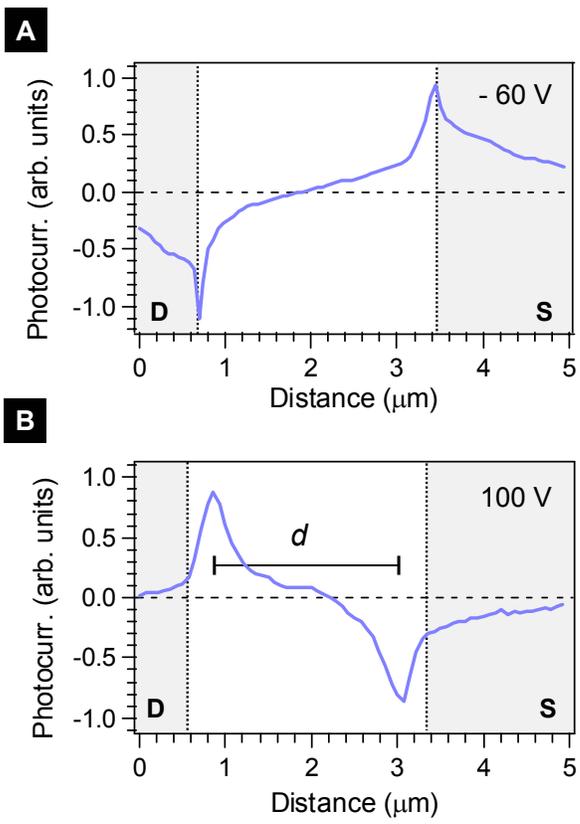

Figure 3, Mueller et al.

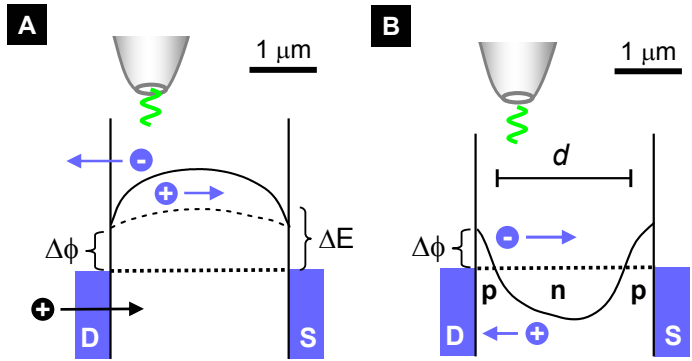

Figure 4, Mueller et al.

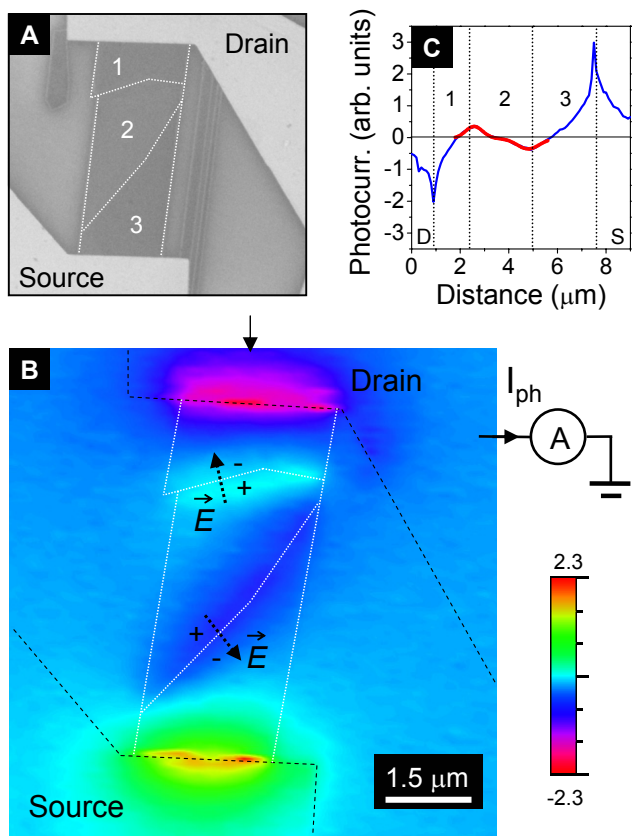

Figure 5, Mueller et al.